\documentclass[reprint,amsmath,amssymb,aps,footinbib]{revtex4-1}
\usepackage{dcolumn}
\usepackage{color}
\usepackage{bm}
\usepackage{xspace}
\usepackage{graphicx} 

\newcommand{\be}{\begin{equation}}
\newcommand{\ee}{\end{equation}}
\newcommand{\bea}{\begin{eqnarray}}
\newcommand{\eea}{\end{eqnarray}}
\newcommand{\beaa}{\begin{eqnarray*}}
\newcommand{\eeaa}{\end{eqnarray*}}
\newcommand{\alphaI}{{$\alpha$-(BEDT-TTF)$_2$I$_3$}\xspace}

\begin{document}

\title{
  Coherent Interlayer Coupling in Quasi-Two-Dimensional Dirac Fermions
  in \alphaI
}

\author{Naoya Tajima}
\email{naoya.tajima@sci.toho-u.ac.jp}
\affiliation{Department of Physics, Toho University, Funabashi, Chiba 274-8510, Japan}
\author{Takao Morinari}
\email{morinari.takao.5s@kyoto-u.ac.jp}
\affiliation{Course of Studies on Materials Science,
  Graduate School of Human and Environmental Studies,
  Kyoto University, Kyoto 606-8501, Japan
}
\author{Yoshitaka Kawasugi}
\affiliation{Department of Physics, Toho University, Funabashi, Chiba 274-8510, Japan}
\author{Ryuhei Oka}
\affiliation{Department of Chemistry, Ehime University, Matsuyama 790-8577, Japan}
\author{Toshio Naito}
\affiliation{Department of Chemistry, Ehime University, Matsuyama 790-8577, Japan}
\author{Reizo Kato}
\affiliation{RIKEN, Hirosawa 2-1, Wako-shi, Saitama 351-0198, Japan}

\date{\today}

\begin{abstract}
  Theoretical and experimental studies have supported
  that the electronic structure of \alphaI under pressure
  is described by two-dimensional Dirac fermions.
  When the interlayer tunneling is coherent,
  the electronic structure of the system becomes three-dimensional,
  and we expect the peak structure to appear in the interlayer resistivity
  under magnetic fields.
  We theoretically and experimentally show that the peak appears
  in the interlayer resistivity at low temperatures
  and high magnetic fields.
  From the experiment, we estimate that the magnitude of the interlayer tunneling
  is $t_1 \sim 1$ meV.
  Our result opens the door to investigating the three-dimensional
  electronic structure of \alphaI.
\end{abstract}

\maketitle

The research on massless Dirac fermion systems is one of the central subjects
in modern condensed matter physics.
In organic conductors,
theoretical and experimental studies support that
\alphaI under high pressure is a quasi-two-dimensional
Dirac fermion system.\cite{Katayama2006,Kajita2014}
(Here, BEDT-TTF is bis(ethylenedithio)tetrathiafulvalene.)
The system consists of alternating layers of
BEDT-TTF molecules and insulating I$_3^-$ anions.\cite{Bender1984}
The band calculation suggested\cite{Katayama2006} that
the electronic structure is described by massless Dirac fermions
under high pressure.
The first band and the second band touch each other
at two points in the first Brillouin zone
where the bands exhibit linear energy dispersion.
There are two more bands, but they are energetically well separated
from the linearly dispersing bands.
Since there is one hole per two molecules,
the HOMO band is 3/4-filled.
As a result, the Fermi energy is located precisely at the Dirac point.
The inter-layer magnetoresistance clearly demonstrates the presence
of the zero-energy Landau level of the Dirac fermions\cite{Osada2008,Tajima2009}
that leads to a negative magnetoresistance.
Furthermore, the phase of the Dirac fermions is confirmed by
the Shubnikov–de Haas oscillation of the hole-doped sample,
where the sample is placed on polyetylene naphthalate substrate.\cite{Tajima2013}

One of the present authors proposed\cite{Morinari2020} that,
upon entering a coherent interlayer tunneling regime,
the system becomes a three-dimensional
Dirac semimetal phase.\cite{Young2012a,Wang2012,Yang2014}
Intriguing transport properties, such as a negative magnetoresistance
associated with chiral anomaly\cite{Xiong2015} and a planar Hall effect,\cite{Burkov2017,Nandy2017}
are expected in the three-dimensional Dirac
semimetal state.\cite{Armitage2018}
In order to investigate this possibility,
we need first to establish the presence of the coherent
interlayer tunneling.
If the interlayer tunneling is coherent,
a peak structure appears in the interlayer magnetoresistance
when the applied magnetic field is almost in the plane.\cite{Hanasaki1998}
Moreover, the width of the peak provides an estimation of
the magnitude of the interlayer tunneling.

In this Letter, we report that the peak appears
in the interlayer magnetoresistance in \alphaI under pressure.
The result suggests that the interlayer tunneling
is coherent in \alphaI at low temperatures.
From the experiment, we etimate that
the interlayer transfer energy is $t_1 \sim 1$ meV.
In the following, we present a theoretical analysis
and the experimental result.

We formulate the interlayer transport
in \alphaI based on a simple model.
In \alphaI, the unit cell is composed
of four BEDT-TTF molecules, A, A', B, and C.
In the conduction plane, molecules A and A' and molecules B and C are stacking along the $a$-axis.
This structure has interlayer tunnelings
between the same molecules
and between molecules A and A' and B and C.\cite{Morinari2020}
We include these interlayer hoppings,
and take the simplest form for the dispersion
of the two-dimensional Dirac fermion.
The energy dispersion is given by
\be
E_{\bm{k}}^{\left(  \pm  \right)}
=  - 2{t_1}\cos \left( c{k_z} \right) \pm \sqrt {{\hbar ^2}v^2
\left( {k_x^2 + k_y^2} \right) + 4t_2^2{{\cos }^2}{\left( c k_z \right) }}.
\label{eq:Ek_pm}
\ee
Here, $v$ is the Fermi velocity in the plane,
$k_x$ and $k_y$ denote the inplane wave numbers,
$k_z$ denotes the wave number in the direction
perpendicular to the plane,
and $c$ is the lattice constant.
The parameter $t_1$ ($t_2$)
describes the interlayer tunneling amplitude
between the same (different) molecules.
\footnote{An explicit form of the Hamiltonian is given in the Supprelemntaly material in Ref.~\onlinecite{Morinari2020}.}
Reflecting that the energy dispersion in the plane has the form
of the Dirac fermions, there are positive and negative energy states.
The energy dispersion of the positive (negative) energy state is
given by $E_{\bm{k}}^{(+)}$ ($E_{\bm{k}}^{(-)}$).
For \alphaI, it is reasonable to assume that $t_1 > t_2$.
The model describes type-II Dirac fermions\cite{Soluyanov2015} in this case,
where the Dirac cones at $(0,0,\pm \pi/2)$ tilt strongly along the $k_z$ axis,
and both the positive and negative energy states cross the Fermi energy.
In two-dimensional Dirac fermion systems,
there are at least two Dirac points in the two-dimensional
Brillouin zone.
Here we consider them separately,
and the result is multipllied by the number of Dirac points.

We calculate the interlayer conductivity
based on the quasiclassical Boltzmann formalism.
Under the magnetic field ${\bm B}$,
the equation of motion of electrons is given by
\be
\hbar \frac{{d{\bm{k}}}}{{dt}} =  - e{{\bm{v}}_{\bm{k}}} \times {\bm{B}},
\ee
where ${\bm v}_{\bm{k}}={\bm v}_{\bm{k}}^{(+)}$
or ${\bm v}_{\bm{k}}={\bm v}_{\bm{k}}^{(-)}$
with
\be
{\bm{v}}_{\bm{k}}^{\left(  \pm  \right)} = \frac{1}{\hbar }\frac{{\partial E_{\bm{k}}^{\left(  \pm  \right)}}}{{\partial {\bm{k}}}}.
\ee
By solving the Boltzmann equation in the relaxation time approximation,
we obtain Chamber's formula\cite{Ashcroft1976}
for the interlayer conductivity,
\be
   {\sigma _{zz}} = \frac{{2{e^2}}}{{{V}}}
   \sum\limits_{{\bm{k}}\left( 0 \right)}
              {\left( { - \frac{{\partial f}}
                  {{\partial {E_{{\bm{k}}\left( 0 \right)}}}}} \right)
                {{\left[ {{{\bm{v}}_{{\bm{k}}\left( 0 \right)}}} \right]}_z }
                \int_{ - \infty }^0 {dt} {{\rm e}^{\frac{t}{\tau }}}
                    {{\left[ {{{\bm{v}}_{{\bm{k}}\left( t \right)}}} \right]}_z }},
\ee
where $f=f(E_{{\bm k}(0)})$ is the Fermi distribution function,
$\tau$ is the relaxation time,
and $V$ is the volume of the system.
The wave vector at time $t=0$ is denoted by ${{\bm k}(0)}$
and the wave vector at time $t$ is denoted by ${{\bm k}(t)}$.
The integral with respect to time $t$ gives
the electron velocity averaged over trajectories on the Fermi surface.
We compute $\sigma_{zz}$ for the positive
energy state with setting
$E_{\bm k}=E_{\bm k}^{(+)}$
and that for the negative energy state
with setting $E_{\bm k}=E_{\bm k}^{(-)}$, separately.
The sum of them gives the total interlayer conductivity.
In the following theoretical analysis, we take $\hbar v/a$,
the unit of energy with $a$ the inplane lattice constant,
and compute $\sigma_{zz}$ at zero temperature.

\begin{figure}[htbp]
  \includegraphics[width=0.8 \linewidth, angle=0]{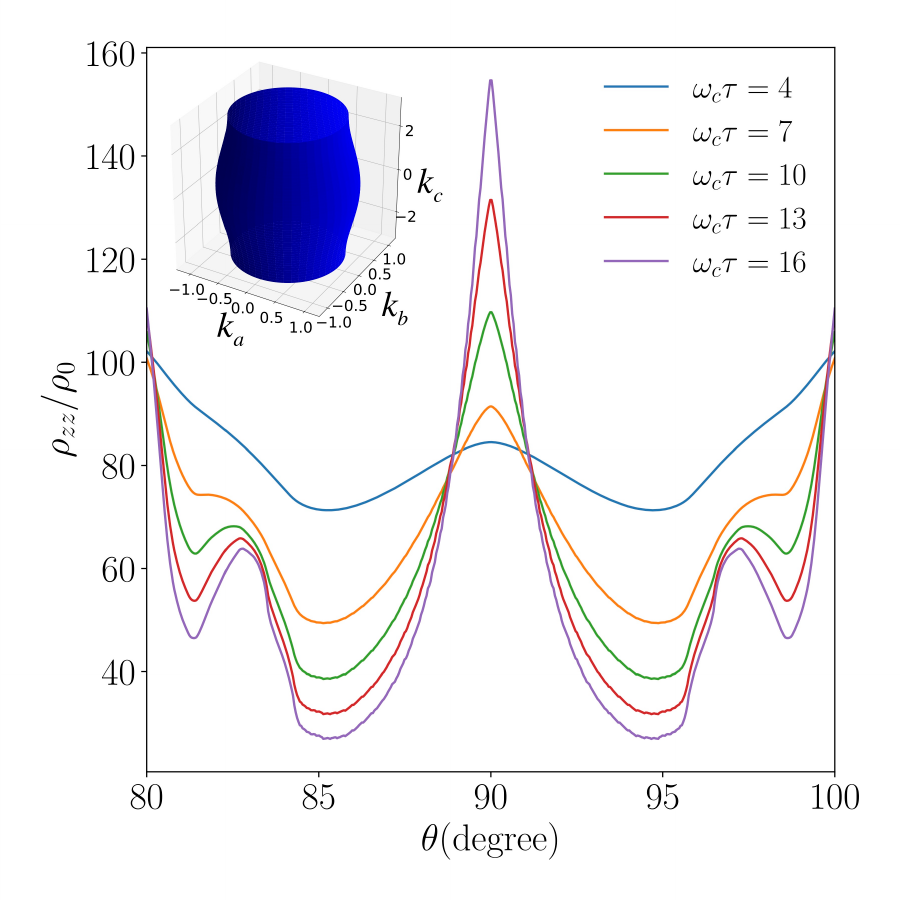}
  \caption{
    \label{fig:numerics:coherence_peak}
    (Color online)
    Dependence of the interlayer resistance
    on the direction of the magnetic field
    for different values of $\omega_c \tau$
    with $\varepsilon_F > t_1, t_2$.
    The interlayer hopping parameters are
    $t_1=0.05$ and $t_2=0.02$ and we take $\varepsilon_F=1$ and $a/c=1$.
    The inset shows the Fermi surface.
  }
\end{figure}

We consider the two cases:
$\varepsilon_F > t_1, t_2$ and $\varepsilon_F < t_1, t_2$.
In the first case, the result
is given in Fig.~\ref{fig:numerics:coherence_peak}
with ${\bm B}=B(\sin \theta, 0, \cos \theta)$.
Here, $\rho_{zz}=1/\sigma_{zz}$ and $\rho_0=1/\sigma_0$
with ${\sigma _0} = {{4{e^2}}}/({{\pi \hbar c}})$.
In the expression of $\sigma_0$, we include the spin and valley
degrees of freedom.
We note that the cyclotron mass\cite{Novoselov2005} is $\varepsilon_F/v^2$,
and the cyclotron frequency is $\omega_c=eBv^2/\varepsilon_F$.
In case of $\omega_c \tau=1$, there is no peak at $\theta=90^{\circ}$,
where the magnetic field is in the plane.
A peak appears at $\theta=90^{\circ}$ for large values of $\omega_c \tau$,
and its height increases as $\omega_c \tau$ increases.
The appearance of the peak at $\theta=90^{\circ}$ is the consequence
of the presence of the small closed orbits.\cite{Hanasaki1998}
Small closed orbits lead to a significant contribution to $\sigma_{zz}$,
and as a consequence, we observe a local minimum (maximum) in $\rho_{zz}$ ($\sigma_{zz}$).
There appears a peak between two local minima in $\rho_{zz}$.
If we neglect the contribution of $t_2$ to ${\bm v}_{\bm k}$,
which is a reasonable assumption for \alphaI,
we have revealed an approximate formula
for the characteristic angle $\theta_c$ for the closed orbits
from the condition, ${\bm v}_{\bm k}\cdot {\bm B}=0$.
That is,
\be
\cot \theta_c  \simeq \frac{{2{t_1}c}}{\hbar v}
\label{eq:theta_c}
\ee
Around $\theta=\pm \theta_c$, $\rho_{zz}$ takes the local minimum.
For the case of $\frac{{2{t_1}c}}{\hbar v}=0.10$,
$\Delta \theta_c \equiv \theta_c - 90^{\circ} \simeq 5.7^{\circ}$.
This estimation agrees with the minima around
$\theta \sim 90 \pm 5^{\circ}$
in Fig.~\ref{fig:numerics:coherence_peak}.
An oscillation for $|\Delta \theta_c| > 6$ is associated
with the Fermi surface effect.\cite{Yamaji1989}
We note that if $t_2 \neq 0$,
there is the chiral anomaly effect
because the system is a three-dimensional Dirac semimetal.
The chiral anomaly plays an important role when the magnetic and electric fields are almost parallel.
However, we may safely neglect it around $\theta \sim 90^{\circ}$
because the magnetic field is almost perpendicular to the electric field.

\begin{figure}[htbp]
  \includegraphics[width=0.8 \linewidth, angle=0]{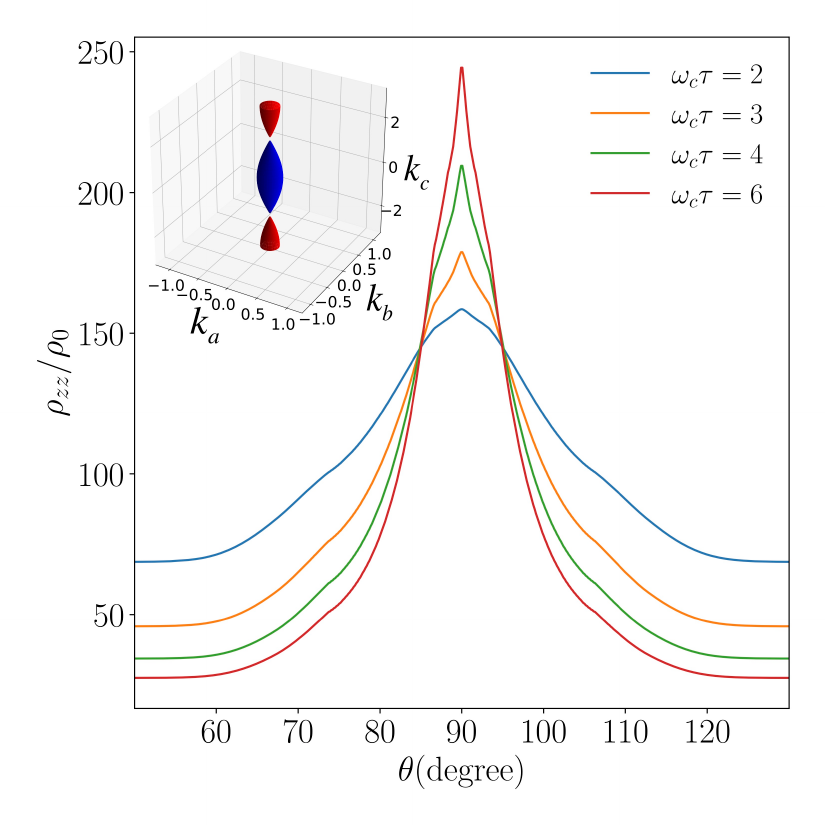}  
  \caption{
    \label{fig:numerics:coherence_peak2}
    (Color online)
    Dependence of the interlayer resistance
    on the direction of the magnetic field
    for different values of $\omega_c \tau$
    with $\varepsilon_F < t_1, t_2$.
    The interlayer hopping parameters are $t_1=0.1$, $t_2=0.05$
    and we take ${\varepsilon}_F=0.02$ and $a/c=1$.
    The inset shows the Fermi surface that consists
    of one electron Fermi surface (blue)
      and two hole Fermi surfaces (red).
  }
\end{figure}

For the case of $\varepsilon_F < t_1, t_2$,
the result is given in Fig.~\ref{fig:numerics:coherence_peak2}.
Similar to Fig.~\ref{fig:numerics:coherence_peak},
we clearly observe the peak at $\theta=90^{\circ}$ for $\omega_c \tau \ge 2$
and its height increases as $\omega_c \tau$ increases.
We find $\Delta \theta_c \simeq 11^{\circ}$ for the parameters
taken for Fig.~\ref{fig:numerics:coherence_peak2}.
Contrary to Fig.~\ref{fig:numerics:coherence_peak},
we do not observe the oscillation associated with the Fermi surface
because the inplane Fermi wave vector is too small
to observe the oscillation in case of $\varepsilon_F = 0.01$.

Now we experimentally demonstrate 
that the interlayer tunneling in $\alpha$-(BEDT-TTF)$_2$I$_3$ under pressure is coherent 
from detecting a peak structure 
in the interlayer magnetoresistance under the magnetic field for $\theta \sim 90^{\circ}$.

The interlayer resistance of $\alpha$-(BEDT-TTF)$_2$I$_3$
under hydrostatic pressures up to 1.7 GPa, 
which were applied using a clamp-type pressure cell made of hard alloy MP35N, 
was measured by a conventional dc method with the electric current of 1 $\mu$ A 
along the $c$-crystal axis normal to the plane. 

\begin{figure}[htbp]
  \includegraphics[width=0.8 \linewidth, angle=0]{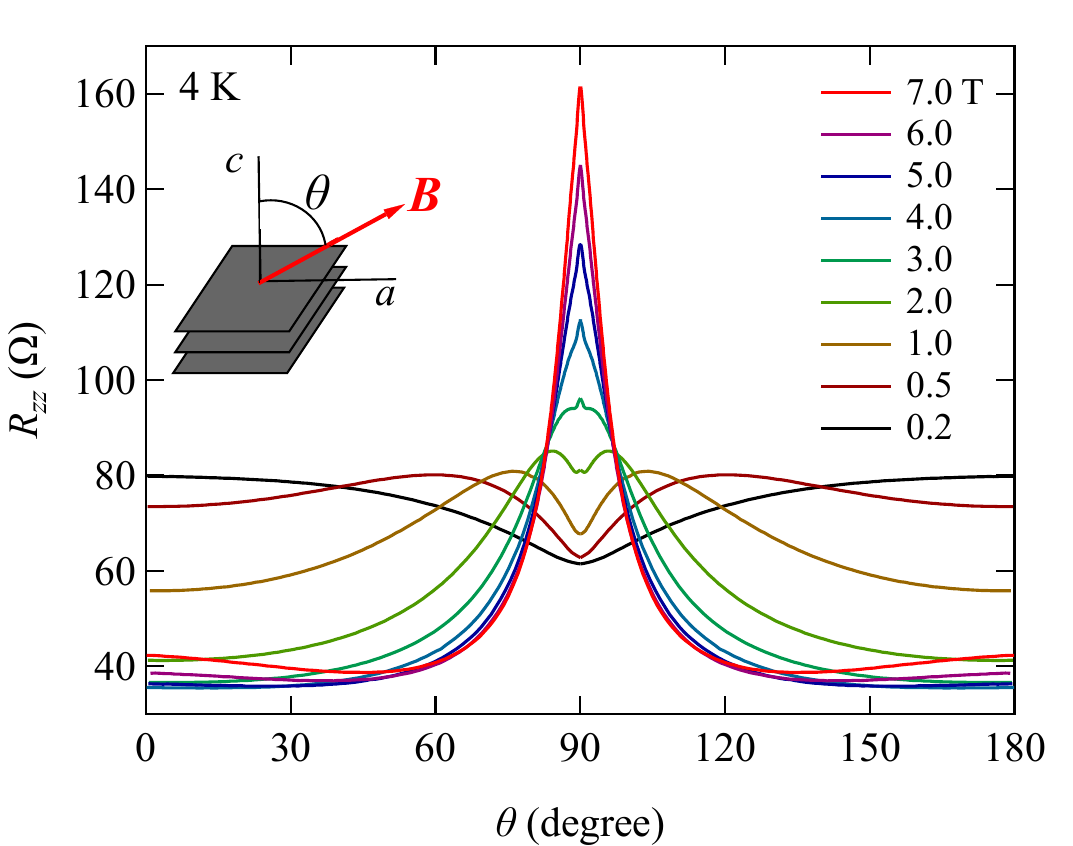}  
  \caption{
    \label{fig:exp:coherence_peak}
    (Color online)
    The angular dependence of the interlayer resistivity for different
    magnetic fields at 4 K.
    The inset shows the definition of $\theta$.    
    The peak at $\theta=90^{\circ}$ is clearly observed
    for $B > 3$ T, while there are no peaks for $B < 2$ T.
  }
\end{figure}

We show angular ($\theta$-) dependent interlayer magnetoresistance
around $\theta=90^{\circ}$ within the $ac$-plane at 4 K in Fig.~\ref{fig:exp:coherence_peak}.
The peak structure of interlayer magnetoresistance
was detected in magnetic fields above 3 T.
The detection of the peak structure
suggests that the interlayer tunneling
is coherent.

Now, we estimate the magnitude of the interlayer hopping. Two main contributions to the interlayer magnetoresistance are coherent interlayer tunneling and the zero-energy Landau level.\cite{Osada2008,Tajima2009} The latter becomes significant around $\theta=0$ and at high magnetic fields. Considering the magnetic field dependence of the contribution from the zero energy Landau level, we estimate
that
$t_1 \sim 1$ meV
from the relationship (\ref{eq:theta_c}) with $v = 5\times 10^4$ m/s and $c=1.75$ nm and $a=1.00$ nm. Here, we take the value of $a$ from the average of the inplane lattice constants.

Note that this interlayer hopping energy is much higher than the Fermi energy. 
The effective carrier density at the lowest temperature indicates
$|\varepsilon_F|/k_{\rm B} < 0.1$ K.
Since we observed peak structures at 4 K,
the order of the magnitudes is $t_1/k_{\rm B} > 4 {\rm K} > |\varepsilon_F|/k_{\rm B}$
where $k_{\rm B}$ is the Boltzmann constant.
To our knowledge, our observation is the first example
that detects the peak structure for $t_1 \gg \varepsilon_F$.
The result is confirmed
by the model calculation result, Fig.~\ref{fig:numerics:coherence_peak2}.

To conclude, we have presented a theoretical analysis of the interlayer conductivity for the quasi-two-dimensional Dirac fermions in \alphaI under magnetic fields and the experimental result supporting the presence of coherence in the interlayer tunneling at low temperatures.
Based on the approximation formula, Eq.~(\ref{eq:theta_c}),
we have demonstrated that $t_1 \sim 1$ meV.

Our result suggests that
the electronic structure
of \alphaI is three-dimensional
at low temperatures and high pressure
and opens the door to investigating
the three-dimensional electronic structure of \alphaI.
In particular, it is critical to
examine whether \alphaI
is a Dirac semimetal at low temperatures.
We will report on it in a future publication.

\begin{acknowledgments}
  This work was supported by JSPS KAKENHI Grant Numbers 20K03870 and 22K03533.
\end{acknowledgments}

\bibliography{../../refs/lib}
\end{document}